# ESTIMATING SOFTWARE RELIABILITY IN MAINTENANCE PHASE THROUGH ANN AND STATISTICS

Ahmad Mateen[1], Muhammad Azeem Akbar[2]
Department of Computer Science, University of Agriculture Faisalabad, Pakistan
ahmedmatin@hotmail.com, azeem.akbar@ymail.com

**ABSTRACT:** *Maintenance is the last and the most critical phase of the software development life cycle. It involves debugging of errors and different types of enhancements which are requested by the user. Software reliability regarding maintenance is the most crucial part as it depends upon the time and cost to correct the errors and make enchantements. It is often felt that software errors or correction takes time to be removed. The maintenance time depends upon the nature of the occurred errors and requested enhancements. In this research work we predict the software reliability in terms of time taken to maintain the errors and enhancements. Artificial Neural Network is used to analyze and predict the software reliability of the maintenance phase. At the end statistical results and proposed neural network results are also compared to make sure that forecasted results are equal to the output results. These results are compared to show that the ANN can understand the relationship between data in a better way. This research work also shows how much it is difficult to understand the nature of maintenance data.*

## 1. INTRODUCTION

There is need to develop and maintain the system continuously and consistently because no process cycle has been developed to engineer secured information system. In software engineering, SDLC facilitates a set of comprehensive task. To retain the system continuously there is a need to revise the tasks such as design, development, test and maintain. Software market competition is increasing day by day all over the world. Customer expectations are higher regarding services and cost [1]. Software maintenance is the critical and the tough part of the SDLC. Software maintenance has totally different meaning and sense as compared to hardware maintenance. Software maintenance requires changes and up gradations. These changes and up gradations might be functional, non-functional and technical [2].

The customer, client or environments of the industries are demanding for these types of changes. Software maintenance can be complex and costly in different circumstances of maintenance. When software is implemented successfully, then maintenance of the software becomes necessary. If the software is not maintained, then it will lead to high cost damages. Information system maintenance costs are a significant expenditure. For some organizations, as much as 80% of their information systems budget is allocated to maintenance activities. In such phase the problem or Enhancements are detected and fixed them so by detecting the caucus it is very expensive. Due to this extra expense the remaining cost low for adaptive maintenance. Eventually, adaptive, perfective, and preventive maintenance activities are effect in increasing the goodwill in a business environment. Due to the adaptive maintenance the end system life expenses for high level correction increased rapidly the need for the maintenance is to check whether system performs well to satisfy the customer requirements. To fix the defects is a general perception behind the concept of maintenance. With the passage of time the system demands different types of the maintenance. It includes the changing of procedures, programs or Documentation for the insurance of correct system and adaption the changing requirements for improving the efficiency of the system

In adaptive maintenance, changes are made in the software according to changes in the software environment. This is done so that software becomes effective. In preventive maintenance, software is preventing from failures and problems that can occur in the future. In perfective maintenance, users demand new or change the requirements of the software. These requirements affect the functionality of the software. Implementation of these new or change requirements is perfective maintenance [3].

Reliability is a non-functional characteristic of any system or software. It means probability of failure free operation in the system in a specified time and in a required environment. The reliability of a system or software becomes low due to the faults and problems in the software. The major issue in the maintenance is cost of the corrections of the occurred problems [4,5,6].

The need of the reliability in the maintenance phase is to lower the faults. When the occurred faults are less in number then automatically cost will be less. Due to the changes or problems in the hardware and the design, the changes have to be made in the software. Errors, ambiguities, carelessness while writing the code, misinterpretation, inadequate testing etc. are the major issues of software failure. So with the passage of time hardware shows the failure characteristics of the software as the hardware are changed according to the challenging and demanding software changes [7,8].

Reliability of the maintenance phase can be predicted using different homogeneous and non-homogenous models [9]. The artificial neural network is the one major approach of artificial intelligence. It is used in different fields such as medical, testing, development, cost prediction, effort estimation, requirement specification, forecasting yield production and weather forecasting [8][10]. ANN is also used to predict the reliability of different phases of SDLC. Such as requirement specification, designing, coding, testing and maintenance phase. If reliability of maintenance phase is predicted then there is lot of chances to control the maintenance cost. So there is need to develop a reliability prediction system to forecast the maintenance phase reliability. It will not only be helpful in improving the reliability of maintenance phase but also helpful in developing a strategy to maintain the system in a cost effective format.

### 1.1. Artificial Neural Network

ANN structure is developed by considering the structure of human brain. Millions of neurons are connected together. The FIGURE 1: is representing the basic structure of 'ANN.

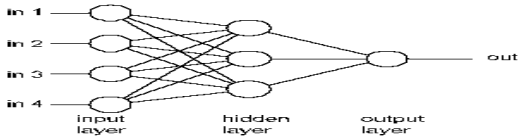

**FIGURE 1: Basic structure of ANN**

According to the model input layer is the first layer, after input layer hidden layer is used with in intermediary layer and output layer is the third layer. The layer contains the different number of neurons. These neurons are interconnected with each other through different nodes. Node is the basic computational element of the artificial neural network. Information is passed out through these neurons. All types of mathematical calculations and consequences are produced by the single neuron which will become the input for the other neurons. All types of processing are done in hidden layer or layers. All the layers are associated with nodes. These nodes or external sources have some associated weight. The data or information is given to the network through input layer. This layer never changes the given values. Each given input value is duplicated and then forwarded to the hidden layer. Hidden layer receives the data from the input layer. Hidden layer, then adjusts some weights of the values and forwarded to the output layer. Then output layer generates results. After it activation function processes the output. ANN is develop and trained for the concerned purpose according to the inputs and outputs. After training it works accordingly [11,12].

Back Propagation Neural Network As the real uniqueness of a network happens in by adjusting the weights. For the solution of the problem the weights are adjusted with particular method. Back propagation is used as a learning algorithm for training the network.

The working of BP is performed in the iterations which is very small steps the examples are applied on the network for training it .the produces output depends upon the current state of the weights. The network produces the output and mean square error is calculated .the occurred error is then propagated backwards through the network. The weights are adjusted to reduce the error the whole process is repeated for each case. This process is repeated until the produced result by the network is acceptable.so that's the position when we can say the network is well learned. The network will never exactly learn the ideal function but it will asymptotically approach the ideal function [17] we have used BPNN algorithm to develop the model.

### 1.2. MAINTENANCE DATA

The data in this research is collected from tracker software; the data is in the form of change log and it is summarized in the excel sheet in the form of input and output. Data is collected from April 2007 to July 2014. The total amount of the data we get is 1 x 56. We build the network model based on the 70% values for training and 15% are used for validation and 15% is reserved for testing purpose. The input of the developed model is accumulated by two types of maintenance such as number of enhancements (e) and correction of failures (f) as shown in eq (1).

$$\text{Input} = e + f \ldots \ldots \ldots \ldots \text{eq (1)}$$

Output is based on the time in terms of days which is taken by the developer to maintain the correction and enhancements as shown in Table 1. Error correction record is actually referred to those failures which demand solution on the spot. Other all types of maintenance such as adaptive, perfective and preventive maintenance are called enhancement maintenance record. The purpose behind all this work is to predict the days to maintain the errors.

**Table 1: Maintenance Data**

| MAINTENANCE | | | TIME (days) | | |
|---|---|---|---|---|---|
| Enhancements (e) | Corrections (f) | Input (X) | Time (e) | Time (f) | Output (Y) |
| 5 | 5 | 10 | 17 | 8 | 25 |
| 11 | 9 | 20 | 23 | 20 | 43 |
| 5 | 8 | 13 | 24 | 13 | 37 |
| 4 | 5 | 9 | 10 | 16 | 26 |

### 1.3. ASSUMPTION RELATED TO THE MODEL

Number of corrections and the enhancements are summed up as the input. Output is consisting of days taken to maintain the correction and enhancements. All the other parameters such as developer's boredom, tiredness, etc. are kept constant or neglected. The occurrence of errors will continous in the reliability issues. Therefore, the maintenance of software will also continue till the software life.

### 1.4. PROPOSED MODEL

The developed neural network is based on the back propagation algorithm. BP algorithm is very flexible in its working capacity. There are three layers in developed model. The first layer is inputs which have one neuron. The input values are taken through this layer. Second layer is hidden layer which have 10 neurons to process the given data. Output layer is the third layer which is consists of one neuron to represent the results. The transfer function, weights and biases of the inputs are controlled by the combination of hidden and output layers. The transfer functions are the algebraic function which may be linear and nonlinear. The connection between the neurons is adjusted by the weights. When the data is presented as the input to the neural network then to develop the pattern of data, weights and biases are adjusted accordingly. It is done so that the desired output can be generated. Weights and biases are adjusted according to the BPN learning rule. After all this adjustments, network is trained to a maximum validated level so that model can perform to produce desired results.

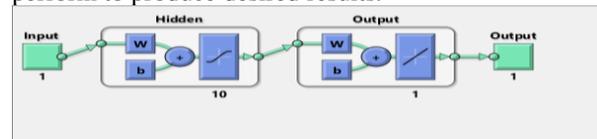

**FIGURE 2: Basic structure of proposed NN**

FIGURE 2: shows the basic structure of the BP neural network which is developed in the MATLAB neural network

toolbox. The structure shows that the one input and output layer while one hidden layer. One neuron is in the input and output layer. Ten neurons are working in the hidden layer of the ANN. Hyperbolic Tangent function is working in hidden layer and in output layer linear activation function is used for training of the model, BP Levenburg Marquardt approximation is used.

## 2. MODEL RESULTS

When enhancement record is as input and days taken to maintain the enhancements are as the output then the results of ANN are shown in fig 3. The dashed line shows the perfect results such as output = targets. There are too many variations in the data sets. When R value is close to 1 then it means the relation between input and target is significant. The R value is 0.66 which shows the correlation between the data values is somehow significant. As other variables are not considered in the data, therefore this regression value is significant in comparison. The line represents the relationship between the values. The points which are far from the dashed line are called outliers. In validation plot the training graph indicates that the fit is not good as mostly data values are outliers.

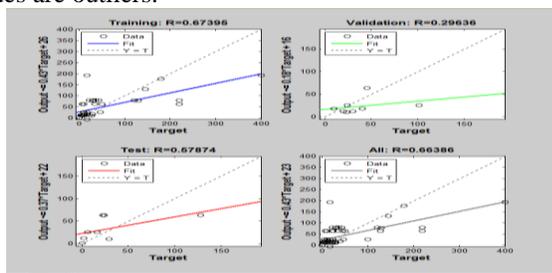

**FIGURE 3a: Regression Plot of Enhancement record**

The FIGURE 3a: describes issue regarding training. Test, validation and train curves are somehow alike. The figure indicates that there is some overfitting as the test curve has increased before the validation curve. Overfitting means there is too much complication in data and other parameters which are not considered in this practice having more influence on data sets. In the performance graph (FIGURE 3b) poor performance depict the overfitting. The performance of the ANN is best at the epoch 3 according to the given data.

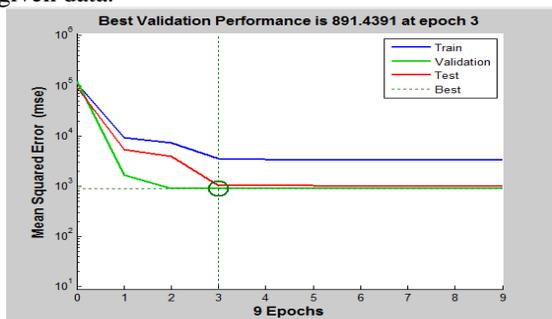

**FIGURE 3b: Performance Plot of ER**

In FIGURE 3c: Histogram shows the outliers in the data set. The rectangles of the histogram show that how much data is continuous. Here bins indicate the equal size of intervals. This result showed that there is no proper continuity between the original variables. When only corrections records is carried out as input then the results of regression plot are shown in figure 4a. The R value is .64 which is significant but the training graph shows that the fit is somehow good.

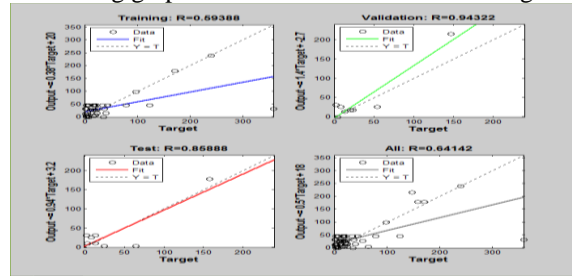

**FIGURE 4a: Regression Plot of Correction Record**

The performance validation of correction record in FIGURE 4b does not indicate overfitting as test curve is significantly decreased as compare to the validation curve. The best validation performance is achieved at epoch 1.

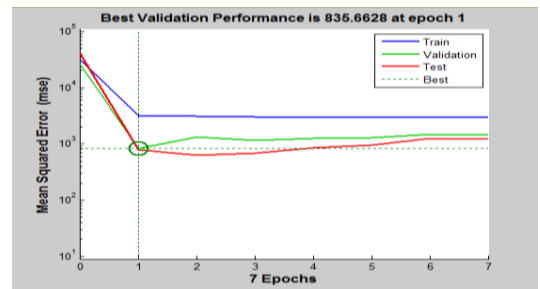

**FIGURE 4b: Performance plot of Correction Record**

In correction record, there is significant continuity between the data variables as shown in figure 4c. As compare to the enhancements record, there is less number of outliers in the data set. If you are required more accurate result then you should retrain the network and for the training you should change the initial weighs and biases of the network, and may this network produce an improved network after retraining.

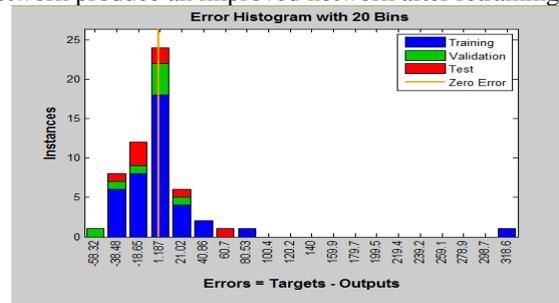

**FIGURE 4c: Histogram plot of Correction Record**

When sum of corrections and enhancements records are taken as input then the regression plot indicates the R value is 0.63. The plot regression command is use to view the regression derived from inputs, targets and outputs. IN figure 5a graph of regression clearly shows the nonlinearity in data. This is actually used to validate the network productivity. The following regression graph show the network outputs with respect to targets for training, validation, and test sets. For a perfect fit, the data should fall along dotted line, where the network outputs are equal to the targets. For this problem, the fit is reasonably good for all data sets, with R values in each case of 0.63.

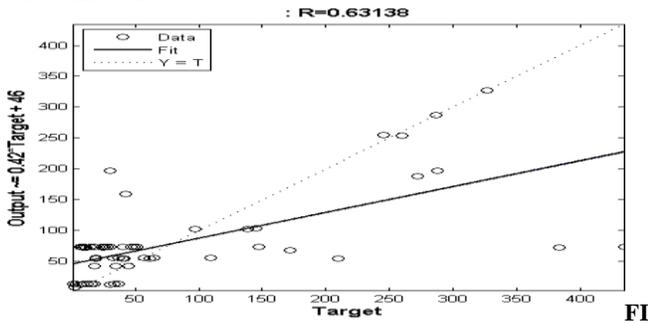

**FIGURE 3a: Regression Plot of Enhancement record**

In figure 5b, the yellow line in the histogram indicates the zero error at 4.25. The blue lines showed the frequency of occurrence of values in specific bin. Bins

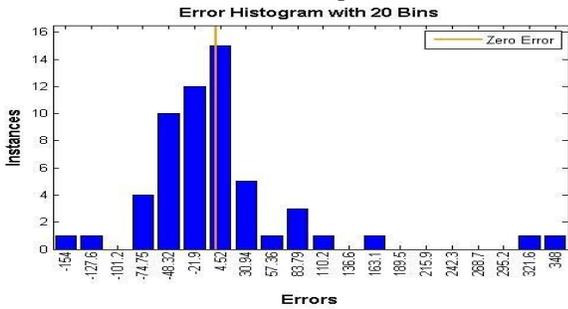

**FIGURE 5a: Regression plot of sum of correction and enhancements**

The below performance validation plot in figure 5c shows that the fit is not good as there is no significant behavior of the three curves. Performance Plot shows the mean square error (MSE) dynamics for all given datasets in logarithmic scale. Training MSE is always decreasing, so its validation and test MSE where I should be interested in is increasing somehow which shows that the output and targets going to significant. The plot shows a good training.

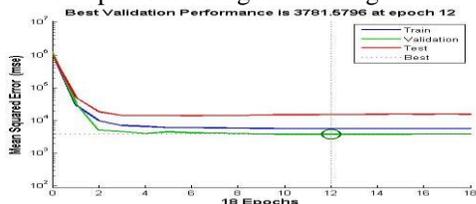

**FIGURE 5b: Histogram plot of sum of correction and enhancements record**

Error corrections and enhancements are different in nature some errors take more time to be maintained but on the other side some errors take less time. It is obvious that complicated errors take more days for maintenance while simple or average type of error takes less time. Therefore, the need is to understand the nature of the errors and categorize in simple, average, complex and very complex categories. A regression analysis is also applied through SPSS (a statistical tool) then the regression plot shows different R value .81 as shown in Table 2 which is highly significant as compare to the developed model. It is not necessary that the result of ANN and the statistical tools must be same. ANN results showed that the data is highly nonlinear and there is need to retrain the model to understand the relationship between the input and output.

The regression graph in figure 6a shows that the data points are somehow close to the regression line. As the number of errors increased or decreased there is no significant variation in days. It means that the ANN can understand the relationship between the input and output more wisely.

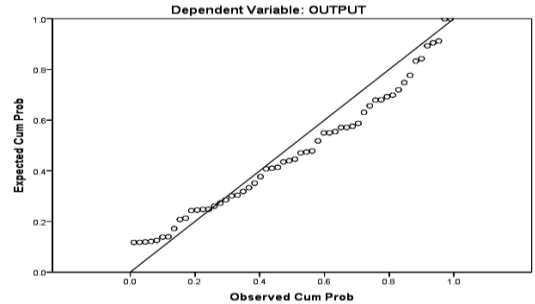

**FIGURE 6a: Regression result through SPSS**

The histogram is indicated the different results from the developed model as shown in figure 6b. As this tool indicate that there is more continuity and less outlier in the data set. The bar height showed that mostly values were fall in that interval. The calculated results through SSPSS are shown in the table .2 .that explain the figure 6a as well as histogram 6b

**Table 2: Regression results through statistical tool (SPSS)**

| Model | R | R Square | Adjusted R Square | Std. Error of the Estimate |
|---|---|---|---|---|
| 1 | .815[a] | 0.664 | 0.651 | 62.944 |

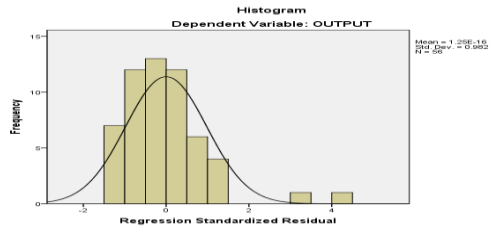

**FIGURE 6b: Histogram plot through SPSS**

### 3.∴.CONCLUSION

Maintenance is the critical phase of the software development life cycle. The results and graph of the neural network indicates that the reliability of software must have to predict by such type of networks. The analysis of ANN and statistical results showed that ANN understands the data complexity and relationship more wisely. Through such type of networks it can be possible to predict that how much can be utilized to maintain any software related problem. We raise the attention through this work that there is need to do a lot of work to predict the software reliability in most authenticated way. Data related to the maintenance must be recorded in this way so that it can be utilized for such type of purposes. If the errors are categorized according to their nature then it may be possible to handle such type of data in more efficient way.